\documentclass[9pt,twocolumn,twoside]{opticajnl}
\journal{opticajournal} 

\setboolean{shortarticle}{true}

\usepackage{xfrac}
\usepackage{multicol}
\usepackage{lineno}

\title{Theory of multicolor soliton microcombs}

\author[1,2,*]{Carlo Silvestri}
\author[1,$\dagger$]{Justin Widjaja}
\author[1]{Austin Lin}
\author[1,2]{C. Martijn de Sterke}
\author[1,2]{Antoine F. J. Runge}

\affil[1]{Institute of Photonics and Optical Science (IPOS), School of Physics, University of Sydney, Sydney, New South Wales 2006, Australia}
\affil[2]{ARC Centre of Excellence for Optical Microcombs for Breakthrough Science (COMBS), School of Physics, University of Sydney, Sydney
2006, Australia}
\affil[$\dagger$]{Current address:{Department of Applied Physics, California Institute of Technology, Pasadena 91125, CA, USA}}

\affil[*]{Corresponding author: carlo.silvestri@sydney.edu.au}

\begin{abstract}
We present a general theory of multicolor soliton microcombs. These frequency combs require specially engineered dispersion and have an optical spectrum consisting of multiple spectral windows, centered at distinct frequencies. Our theory is based on a multiple-scale approach applied to the Lugiato-Lefever equation, and provides a framework to investigate different pumping configurations. For multi-frequency pumping, we predict a progressively lower pumping threshold as the number of spectral windows increases due to an enhancement of the effective nonlinear parameter. However, multi-frequency pumping is not a prerequisite for the formation of these combs and can emerge even with a single driving field. Our theoretical predictions are in excellent agreement with numerical simulations.
\end{abstract}

\setboolean{displaycopyright}{false} 

\begin{document}
\bigskip

\maketitle
\noindent 
Over the past decade, the development of optical frequency comb (OFC) sources based on optically pumped Kerr ring microresonators has been an intensely active research area~\cite{Kippenberg2018}. This is due to their suitability for on-chip integration and their relevance across various application fields, including telecommunications, sensing, optical clocks, high-precision spectroscopy, astronomy, and metrology~\cite{Kippenberg2018, Bao2021, Marin-Palomo2017, Newman:19, Chang2022}. These {\sl microcombs} correspond to solitons propagating within the microresonator cavity, forming through a balance of dispersion and Kerr nonlinearity, along with parametric gain and cavity loss~\cite{Kippenberg2018, Pasquazi2018}. 

Two key challenges in the development of microcombs are the direct stabilization of the comb offset frequency, and the generation of combs broad enough to extend beyond the near-infrared spectral range, with particularly promising implications for spectroscopy. One recent approach developed to address both these challenges is the generation of multicolor soliton microcombs~\cite{Luo:16}. These combs are characterized by a spectrum that comprises multiple spectral windows, each with a characteristic hyperbolic secant envelope centered at distinct frequencies. In time, they correspond to a single soliton consisting of a rapidly varying carrier modulated by the pulse-like temporal profile of a single-color soliton~\cite{Tam_2020}. These multicolor soliton microcombs form in the presence of multiple regions of anomalous dispersion, linking their practical realization to the capacity for precise dispersion engineering within microresonators. While these states have been demonstrated in mode-locked fiber lasers~\cite{Lourdesamy2022, Mao_2021, Widjaja_2024}, they are yet to be observed in coherently driven passive resonators, though designs of dispersion-engineered ring microresonators supporting two-color solitons have been reported~\cite{Moille2018}. However, despite recent numerical investigations of two-color soliton microcombs~\cite{Matsko2023}, a complete theoretical description of multicolor soliton microcombs is yet to be reported.

\begin{figure*} [t!]
\centering
   \includegraphics[width=0.93\textwidth]{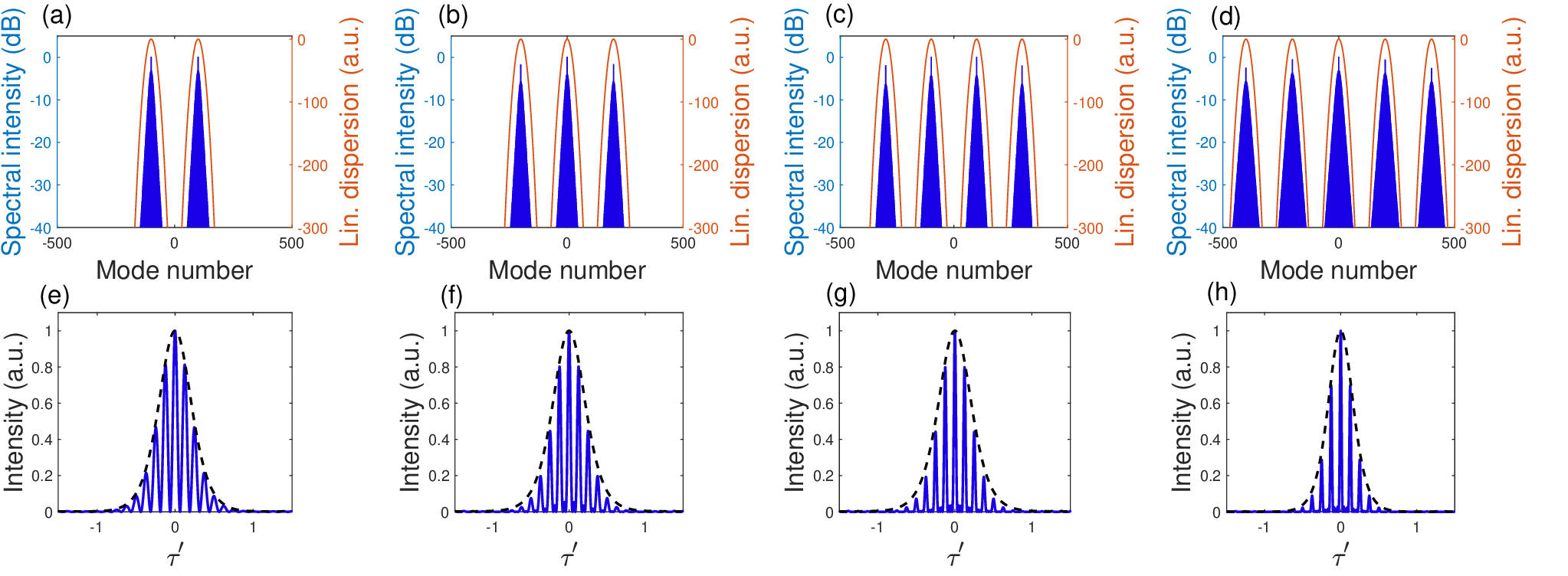}
   \vskip -2mm
   \caption{Multicolor soliton microcombs obtained by numerically solving the normalized LLE \cite{Coen2013} under the optical pumping condition $\sigma_J=1$, for $J=2, 3, 4, 5$. These regimes are achieved by positioning $J$ pumps, each at one of the resonance frequencies of the dispersion peaks, and normalizing their amplitudes such that $h(\tau) = c(\tau)$. (a)-(d): optical spectrum (blue line) and linear dispersion relation (orange line). (e)-(h): intracavity intensity profile of the multicolor soliton (blue line) compared to the corresponding single soliton (black dashed line) obtained for the same detuning and pump power.} \label{fig1}
\end{figure*}

In this letter, we present a general theory of multicolor soliton microcombs, using the multiple-scale method~\cite{Dodd_1982} applied to the Lugiato-Lefever equation (LLE)~\cite{Lugiato1987, Pasquazi2018}. The derivation builds on a similar approach used for the nonlinear Schrödinger equation in Refs~\cite{Lourdesamy2022, Lourdesamy:23}, adapted here for the dissipative nature of the LLE. We consider $J$ equally spaced spectral regions with an angular frequency spacing $\Delta\omega$ and identical dispersion $\Bar{\beta}_2$ (see orange curves in Fig.~\ref{fig1}(a)-(d)). The corresponding spectral profile of the electric field consists of $J$ distinct spectral comb windows. In the time domain, the electric field comprises a rapidly varying carrier and a {\sl meta-envelope} (see Fig.~\ref{fig1}(e)-(h)). Our results show that for each $J$, the meta-envelope satisfies an effective LLE characterized by the dispersion coefficient $\Bar{\beta}_2$, a nonlinear enhancement factor $\varepsilon_J$, and a pump reduction factor $\sigma_J$. While $\varepsilon_J$ depends solely on the carrier properties and is independent of the pump configuration, $\sigma_J$ varies with the number of pumps and their relative intensities. Our theory also suggests the possibility of generating multicolor soliton microcombs with a single pump, where the required pump threshold remains independent of $ J $ and is only slightly higher than the threshold for standard single-color solitons, consistent with numerical simulations. This theory provides a unified framework to analyze diverse pumping configurations and predicts the pump threshold for each scenario. It confirms the reduction of the pump threshold for dichromatic pumping and predicts even greater reductions when $J$ and the number of pumps exceed two~\cite{Matsko2023}. We expect these findings to have significant implications for frequency comb self-referencing and spectroscopy~\cite{Kippenberg2018, Moille2018}.

We start from the generalized LLE, which describes the evolution of the electric field envelope $E$ in a ring resonator~\cite{Lugiato1987,Pasquazi2018}:
\begin{align}
\frac{\partial E}{\partial t}&= \frac{1}{t_R}\left[-\alpha_0-i\delta_0+iL\left(\sum_{n=2,4,...}^{N}(i)^n \frac{\beta_n}{n!} \frac{\partial^n }{\partial \tau^n}\right)+i\gamma L|E|^2\right]E \nonumber\\&+\frac{\sqrt{\theta_0}}{t_R}E_{in}(\tau),\label{LLE1}
\end{align}
where $t$ is the slow time, $\tau$ is the fast time, $t_R$ is the cavity roundtrip time, $\alpha_0$ is the loss coefficient, $\delta_0$ is the detuning, $L$ is the cavity length, $\beta_n$ is the $n$-th order dispersion coefficient, $\gamma$ is the Kerr nonlinear coefficient, and $\theta_0$ is the coupling coefficient for the injected field $E_{in}\left(\tau\right)$. We note that $E_{in}\left(\tau\right)$ is in general a function of $\tau$ to account for case of multi-frequency  pumps. We express $E_{in}(\tau)=E_0h(\tau)$, where $E_0$ is the pump amplitude, and $h(\tau)$ represents the {\sl pump configuration}, and is normalized such that $\langle h^2 \rangle=1$. Here $\langle.\rangle$ indicates the average over a period $2\pi/\Delta\omega$. The dispersion coefficients in \eqref{LLE1} are chosen to generate a dispersion relation $\beta=\beta\left(\omega\right)$ with $J$ equally spaced maxima, with angular frequency spacing $\Delta\omega$. At each of these maxima $\beta=0$, and the curvature is $\bar{\beta}_2$ (see Fig.~\ref{fig1}(a)-(d)). We then write the electric field envelope as~\cite{Tam_2020, Lourdesamy2022}:
\begin{align}
     E\left(\tau, t\right)=\frac{1}{\textit{N}}\varepsilon f\left(\varepsilon\tau,\varepsilon^2t\right)c(\tau)=\frac{1}{\textit{N}}\varepsilon f\left(\varepsilon\tau,\varepsilon^2t\right)\sum_{j}p_je^{-ij\Delta\omega\tau}, \label{electicfield}
\end{align}
where $f$ is the meta-envelope, $c\left(\tau\right)$ is the rapidly varying carrier, which is not known a priori, and the $p_j$ are the amplitudes of its frequency components. Furthermore, $\epsilon$ is a smallness parameter and $N$ is a normalization factor. The additional assumptions of the multi-scale approach, characteristic of the LLE, concern the terms pump, losses and detuning, which are considered to be of the third order in $\epsilon$:
\begin{eqnarray}
\frac{\sqrt{\theta_0}}{t_R}E_{in}\left(\tau\right)&=&\varepsilon^3\frac{\sqrt{\theta}}{t_R}E_{0}h\left(\tau\right),\label{PumpMS}\\
\frac{\alpha_0}{t_R}E(t,\tau)&=&\varepsilon^3 \frac{\alpha}{t_R} \frac{1}{\textit{N}} f\left(\varepsilon\tau,\varepsilon^2t\right)c(\tau).\label{alphaMS}
\end{eqnarray}
An equation of the same form as \eqref{alphaMS} can be written for the detuning, replacing $\alpha_0$ and $\alpha$ with $\delta_0$ and $\delta$, respectively. We substitute Eqs.~(\ref{electicfield})-(\ref{alphaMS}) into \eqref{LLE1} and separate the terms based on their orders in $\epsilon$. The equations at orders $\epsilon$ and $\epsilon^2$ are trivially satisfied \cite{Lourdesamy2022}. At order $\epsilon^3$, we obtain the relevant equation governing the meta-envelope dynamics:
\begin{align}
        \frac{\partial f}{\partial t} =\frac{1}{t_R}\!\!\left[-\left(\alpha+i\delta\right)f-\frac{iL\bar{\beta}_2}{2}\frac{\partial^2 f}{\partial \tau^2}\!
    +i\varepsilon_J\gamma L|f|^2f+\sigma_J\sqrt{\theta}E_{0}\right]\!\!.\label{LLE_meta_envelope} 
\end{align}
and we find the carrier $c(\tau)$, and hence the amplitudes of its frequency components $p_j$, by a self-consistency argument \cite{Lourdesamy2022}. For example, for $J=3$ we find that $p_0=1$ and $p_{\pm 1}=\sqrt{2/3}$.
Thus, the evolution of $f$ is governed by a LLE with second-order dispersion $\bar{\beta}_2$, effective nonlinearity $\varepsilon_J\gamma$, and effective pump amplitude $\sigma_J E_0$. The coefficients $\varepsilon_J$ and $\sigma_J$ take the form
%
\begin{eqnarray}
        \varepsilon_J=\frac{\langle c\left(\tau\right)^4\rangle}{\langle c\left(\tau\right)^2\rangle^2}, \label{epsiloJ}\quad
        \sigma_J=\frac{\langle c\left(\tau\right)h\left(\tau\right)\rangle}{\langle c\left(\tau\right)^2\rangle}. \label{sigmaJ}
\end{eqnarray}
Here, $\varepsilon_J$ is the {\sl nonlinear enhancement factor}, which also enters in the conservative case \cite{Lourdesamy2022}, and depends solely on the carrier $c\left(\tau\right)$. The coefficient $\sigma_J$, characteristic of the multi-scale approach applied to the LLE, serves as an {\sl efficiency factor} for the pump, and is determined by the degree to which the pump configuration $h(\tau)$ matches the carrier $c(\tau)$. The maximum value that $\sigma_J$ can attain is unity, when $h(\tau) = c(\tau)$. Such a scenario arises when the system features $J$ pumps, each positioned at one of the $J$ maxima of the dispersion relation, and the relative pump amplitudes match the relative carrier amplitudes.

The theory developed thus far requires all physical quantities to be explicitly defined, in order to assign an order in $\epsilon$ in the multi-scale approach. For this reason, we adopted the non normalized formulation of the LLE provided in \eqref{LLE1}. However, in \cite{Coen2013}, a normalized form of the LLE was introduced. This formulation relies on a reduced set of parameters, making it more convenient for numerical simulations and facilitating physical insight. For this reason, the simulations presented in this work have been obtained by numerically solving the normalized version of the LLE \cite{Coen2013,Pasquazi2018}. In this formalism the effective LLE for the normalized meta-envelope $f^\prime$ reads:
\begin{eqnarray}
            \frac{\partial f^{\prime}}{\partial t^{\prime}} =\left[-1-i\Delta+i\varepsilon_J |f^{\prime}|^2+i\bar{d}_2 \frac{\partial^2 }{\partial {\tau^{\prime}}^2}\right]f^{\prime}+\sigma_JS_0\label{LLE_envelope_scaled} 
\end{eqnarray}
where $\Delta$, $S_0$ and $\bar{d}_2$ are respectively the normalized detuning, pump amplitude and dispersion coefficient, while $t^{\prime}$ and $\tau^\prime$ are respectively the scaled slow time and fast time. We define the average pump power  $X =\langle (S_0 h\left(\tau^{\prime}\right))^2\rangle$.


\begin{figure} [t!]
   \centering
   \includegraphics[width=0.43\textwidth]{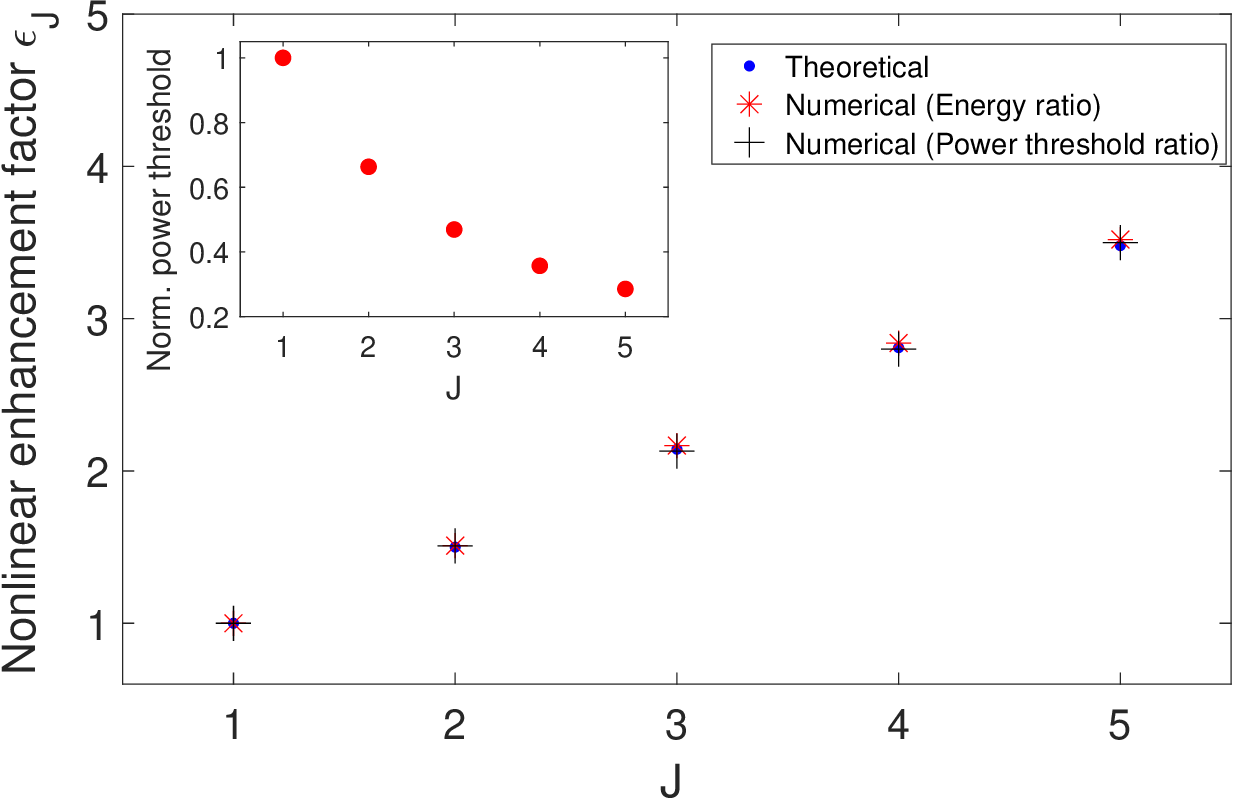}
     \vskip -2mm
   \caption{Theoretical and numerical nonlinear enhancement factor $\varepsilon_J$ as a function of $J$ assuming optimal pumping $\sigma_J=1$ (main panel). The inset shows the power thresholds normalized to the $J=1$ case (red dots).} \label{fig2}
\end{figure}

We first consider a pump such that $h(\tau) = c(\tau)$, corresponding to the highest pump efficiency $\sigma_J = 1$. Figure~\ref{fig1} shows the optical spectrum and intracavity intensity profile for multicolor soliton microcombs generated in numerical simulations under this condition, with a dispersion relation exhibiting $J=2,\cdots,5$ peaks. The spectra are divided into $J$ windows, each centered on one of the dispersion peak frequencies. Consequently, the cases shown in Fig.~\ref{fig1}(b)-(d) provide a generalization of the two-color case of Fig.~\ref{fig1}(a), typically discussed in the literature \cite{Matsko2023, Moille2018}. The temporal intensities (Figs.~\ref{fig1}(e)-(h)) consist of 
the rapidly varying carrier modulated by the envelope which is a solution to \eqref{LLE_envelope_scaled}. As $J$ increases, the carrier exhibits increasingly taller and narrower features, which is  linked to the  increased effective nonlinearity \cite{Lourdesamy2022}, quantified by the nonlinear enhancement factor $\varepsilon_J$. The values of $\varepsilon_J$ predicted by the theory are the same as those in the conservative case \cite{Lourdesamy2022} and increase with $J$ (see blue dots in Fig.~\ref{fig2}). The nonlinear enhancement indicates that, for the same detuning and pump power, it is possible to achieve the same envelope with lower energy $E_\mathrm{J}$ compared to the energy $E_1$ of the conventional single peak soliton case, represented by the dotted black line in the panels of Fig.~\ref{fig1}(e)-(h). Consequently, we can estimate $\varepsilon_J$ as follows:
\begin{align}
    \varepsilon_J=\frac{E_\mathrm{1}-E_\mathrm{bg}}{E_\mathrm{J}-E_\mathrm{bg}} \label{epsmethod1}
\end{align}
where $E_\mathrm{bg}$ is the energy of the background filling the cavity. Numerical values of $\varepsilon_J$ were determined through \eqref{epsmethod1} by numerically solving the LLE using a fourth-order split-step Fourier method \cite{AMIRANASHVILI}. They are indicated with a red marker in the main panel of Fig.~\ref{fig2}, and are in excellent agreement with the theoretical predictions (blue dots). However, since nonlinear enhancement reduces the energy required to trigger soliton formation, it must also influence the minimum pump power at which these regimes can be observe. We, denot thesed as $X^{J,M}_{\text{tr}}$, where $J$ and $M$ are the numbers of dispersion peaks and pump frequency components, respectively. In the inset of Fig.~\ref{fig2}, we present the ratio $X^{J,J}_\text{th}/X^{1,1}_\text{th}$, i.e. the threshold pump power $X^{J,J}_\text{th}$, normalized to the single-soliton threshold $X^{1,1}_\text{th}$, as a function of $J$. The detuning is fixed at $\Delta=12$. We observe a decreasing trend in the threshold pump power with increasing $J$. Since $\sigma_J=1$ for all the regimes considered, the effects of $\sigma_J$ and $\varepsilon_J$ are decoupled. This allows the progressive reduction in the threshold to be attributed specifically to the increasing nonlinearity with $J$. In fact, calculating the reciprocal of the normalized pump threshold provides an estimate of $\varepsilon_J$ (black crosses in the main panel of Fig.~\ref{fig2}) and are in strong agreement with the theory.

\begin{figure} [t!]
   \centering
   \includegraphics[width=0.5\textwidth]{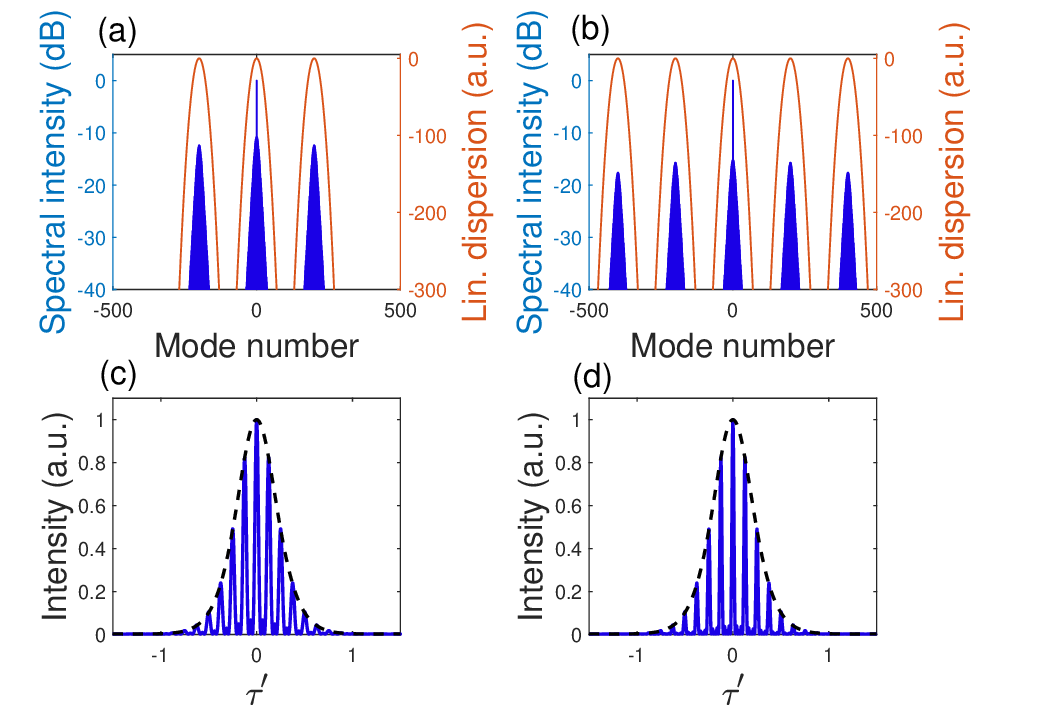}
   \vskip -2mm
   \caption{Spectra (a)-(b) and intensity profile (c)-(d) of multicolor soliton microcombs numerically obtained with a single central pump for $J=3$ (left) and $J=5$ (right).} \label{fig3}
\end{figure}

\begin{figure} [t!]
   \centering
   \includegraphics[width=0.49\textwidth]{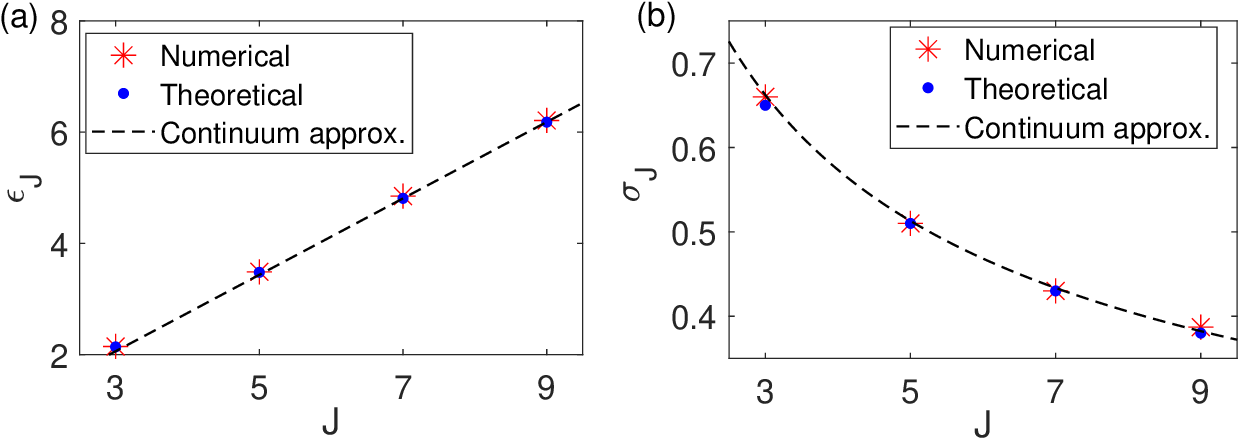}
   \vskip -2mm
   \caption{Single central pump case: theoretical (blue markers) and simulation-estimated (red markers) values of $\varepsilon_J$ (a) and $\sigma_J$ (b) as a function of the number of dispersion peaks $J$. The dashed lines represent the continuum approximation \cite{Lourdesamy:23}.} \label{fig4}
\end{figure}

An additional intriguing prediction of the developed theory is that a single optical pump, operating at the central dispersion peak frequency, can generate a multicolor soliton. This occurs in the presence of $J$ dispersion peaks, where $J$ is odd. This prediction is confirmed by our simulations and in agreement with previous numerical studies \cite{Luo:16}. Examples, generated with a single pump, are illustrated in Fig.~\ref{fig3} for $J=3$ and $J=5$. These solitons have spectra and intensity profiles with the same characteristics as those in Fig.~\ref{fig1}, for the corresponding $J$ values. Consequently, they also present the same nonlinear enhancement factor values. We estimate $\varepsilon_J$ from \eqref{epsmethod1} and find also in this case excellent agreement with the theoretical predictions (see Fig.~\ref{fig4}(a)). Simulations show that these solitons form even if the frequency mismatch between pump and dispersion central peak is as large as $4\times$ the free spectral range.

While the nonlinear enhancement factor remains the same as in the optimal pumping case in Fig.~\ref{fig1}, the theory predicts $\sigma_J<1$ in the presence of a single central pump, i.e. a lower pump efficiency. As an example we consider $J=3$, for which the normalized carrier is $
c\left(\tau\right) = \sqrt{\sfrac{3}{7}}\left[1 + 2\sqrt{\sfrac{2}{3}~}\cos\left(\Delta\omega\tau\right)\right]$. For a single central pump $h\left(\tau\right) = 1$ and using \eqref{sigmaJ}, we find $\sigma_3 = \sqrt{\sfrac{3}{7}~}\approx0.655$, with associated power efficiency ${\sigma_3}^2 = \sfrac{3}{7}$.

To confirm this value through numerical simulations and to further understand the role of $\sigma_J$, we compare two relevant cases: single pump, and three pumps normalized such that $h\left(\tau\right) = c\left(\tau\right)$. As mentioned, the value of $\varepsilon_3$ remains the same in both cases, specifically $\varepsilon_3 = \sfrac{15}{7}$ \cite{Lourdesamy2022}. This decouples the effects of $\varepsilon_3$ and $\sigma_3$, enabling us to estimate $\sigma_3$ from numerical simulations. For a fixed detuning $\Delta = 12$, we determine the minimum pump power required to generate multicolor solitons in both cases. With a single pump, we find $X^{3,1}_\mathrm{th} \approx 10.6$, whereas with three normalized pumps, we obtain $X^{3,3}_\mathrm{th} \approx 4.5$. Their ratio ${X^{3,3}_\mathrm{th}}/{X^{3,1}_\mathrm{th}} = 0.425$ aligns closely with the theoretical prediction ${\sigma_3}^2 = \sfrac{3}{7} \approx 0.429$. For other values of $J$ we find similar agreement between numerical estimates and theoretical predictions (see in Fig.~\ref{fig4}(b)). Thus, for a fixed nonlinear enhancement factor, we find that the power threshold is proportional to $1/\sigma_J$. 

We further compared the numerical and theoretical values of $\varepsilon_J$ and $\sigma_J$ with their continuum approximations (see dashed lines in Fig.~\ref{fig4}) as defined in Ref.~\cite{Lourdesamy:23}. Under this approximation, $\varepsilon_J = c_1 J$, where $c_1 = 0.687$ \cite{Lourdesamy:23}. It can be shown that for a single pump  $\sigma_J^2 = c_2/J$, where $c_2 = 1.316$ (see the Supplementary Material). Although this approximation is rigorously valid only for large $J$, it provides a good estimate of the theoretical values for $J$ as small as $J=3$ (see Fig.~\ref{fig4}). Within the continuum approximation $\varepsilon_J \sigma_J^2 = c_1 c_2 = 0.904$, i.e. this product is independent of $J$. This implies that the power threshold $X^{J,1}_{\text{th}}$ for $J$ dispersion peaks and a single central pump is also independent of $J$. The increased enhancement of the nonlinearity evidently cancels the decrease in the efficiency factor. We can see this as follows: the nonlinear enhancement factor can be estimated as $\varepsilon_J = X^{1,1}_{\text{th}} / X^{J,J}_{\text{th}}$, while the power efficiency for a single pump is $\sigma_J^2 = X^{J,J}_{\text{th}} / X^{J,1}_{\text{th}}$. Combining these, we find $
X^{J,1}_{\text{th}} = X^{1,1}_{\text{th}}/(\varepsilon_J \sigma_J^2)$. Since $\varepsilon_J \sigma_J^2$ is independent of $J$, $X^{J,1}_{\text{th}}$ is also independent of $J$. The value $\varepsilon_J \sigma_J^2 = 0.904$ implies that $X^{J,1}_{\text{th}}$ is only about 10\% higher than the power threshold for the conventional single-color soliton case. For a fixed detuning, therefore, the experimental observation of these multicolor microcombs with a single pump would be feasible at nearly the same pump power levels as required for single-color solitons. We tested this theoretical prediction by numerically estimating the product $\varepsilon_J \sigma_J^2$ through the retrieval of power thresholds from simulations. 

    \begin{figure} [t!]
   \centering
   \includegraphics[width=0.44\textwidth]{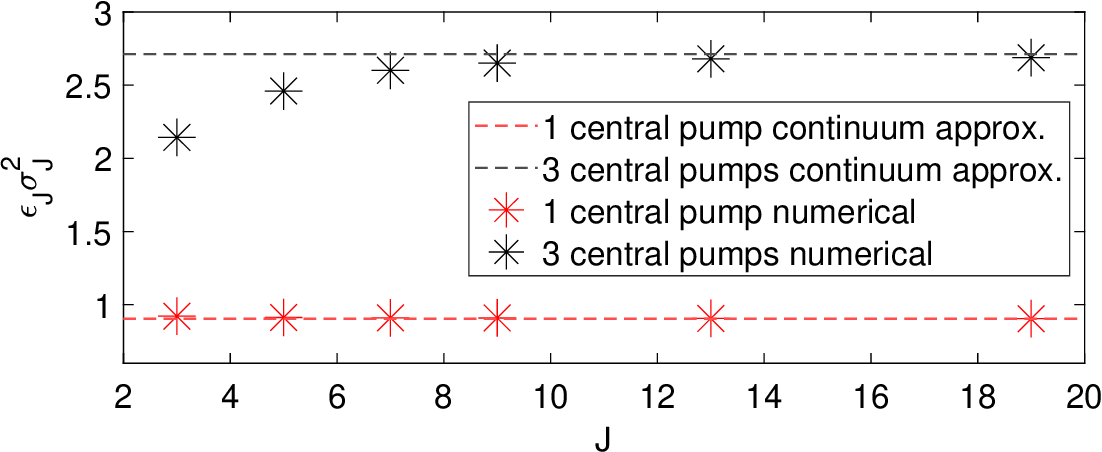}
   \vskip -2mm
   \caption{Continuum approximation (dashed) and numerical estimates (markers) of $\varepsilon_J \sigma_J^2$ for different $J$ for single central pump (red) and three central pump case (black).
} \label{fig5}
\end{figure}

Figure~\ref{fig5} shows that the numerical estimates (red markers) closely match the continuum approximation (red dashed line) for all values of $J$, except perhaps for small $J$, as expected for a theory valid as $J\rightarrow\infty$. We also examined the case of three pumps, normalized such that their amplitudes match those of the three central frequency components, $p_0$ and $p_{\pm1}$, for each $J$. Under the continuum approximation, we have $\varepsilon_J\sigma_J^2 = 3c_1c_2 = 2.712$ (black dashed line in Fig.~\ref{fig5}).  Thus, transitioning from a single pump to three pumps reduces the power threshold by a factor three. Notably, as $J$ increases, the numerical estimates approach the value $3c_1c_2$ (black markers in Fig.~\ref{fig5}).

In conclusion, we developed a theory for multicolor soliton microcombs, identifying two key interplaying physical effects. The first is an enhancement of the nonlinearity, which reduces the pumping threshold for comb formation. The second is an effective pump efficiency, determined by the number of pump frequency components and their relative amplitudes. Our theory also predicts the formation of multicolor microcombs with a single injected field for the same pumping levels as conventional single-color solitons. All our predictions are in excellent agreement with numerical simulations. While we discussed only the case in which the pumping is symmetric, our simulations show that asymmetric pumping leads to more complicated behavior. Our results could enable the observation of comb spectra in spectral regions far from the pump frequency. 

\noindent\textbf{Funding}. Australian Research Council (CE230100006, DE220100509,
DP230102200).\\
\medskip\noindent{\bf\large Acknowledgments} The authors thank Taj Astill, Oliver Sacks and Noah Scollard for early numerical investigations.\\ 
\textbf{Disclosures}. The authors declare no conflicts of interest.\\
\textbf{Data availability}. Data underlying the results presented in this Letter are not publicly available but may be obtained from the authors upon reasonable
request.
\bibliography{sample}



\end{document}